\documentclass{article}
\usepackage{url}
\usepackage{graphicx}

\newcommand{\etal}{et~al.}

\begin{document}

\title{Efficient Batch Update of Unique Identifiers in a Distributed
  Hash Table for Resources in a Mobile Host}
\author{Yoo Chung}
\maketitle

\begin{abstract}
  Resources in a distributed system can be identified using
  identifiers based on random numbers.  When using a distributed hash
  table to resolve such identifiers to network locations, the
  straightforward approach is to store the network location directly
  in the hash table entry associated with an identifier.  When a
  mobile host contains a large number of resources, this requires that
  all of the associated hash table entries must be updated when its
  network address changes.

  We propose an alternative approach where we store a host identifier
  in the entry associated with a resource identifier and the actual
  network address of the host in a separate host entry.  This can
  drastically reduce the time required for updating the distributed
  hash table when a mobile host changes its network address.  We also
  investigate under which circumstances our approach should or should
  not be used.  We evaluate and confirm the usefulness of our approach
  with experiments run on top of OpenDHT.
\end{abstract}

\section{Introduction}
\label{sec:intro}

A distributed system needs a way to identify computing resources that
it uses.  A common way to identify them is to use a
URL~\cite{rfc1738}, which comprises a network address for the host and
a path within the host.\footnote{In
  \texttt{http://example.invalid/1/2/3/}, for example, the network
  address for the host would be \texttt{example.invalid} while
  \texttt{/1/2/3/} would be the path within the host.}  However, some
authors have argued that resource identifiers should contain little or
no information, since otherwise an identifier would become invalid
whenever there are changes in network locations, storage locations,
naming policy, organization,
etc.~\cite{w:cooluri,steen:comm1998,walfish:nsdi2004}

Using random numbers for resource identification avoids including such
information in the identifier.  It also makes it easy to allocate
identifiers without having to manage the identifier space carefully,
since identifier duplication is virtually impossible with a large
enough identifier space.\footnote{Given a 160-bit identifier space,
  the probability that even a single duplicate identifier would be
  generated over a 100-year time period with a billion identifiers
  being generated per second is about $10^{-12}$.}  However, unique
identifiers based on random numbers need to be resolved to actual
network locations, which are often composed of a host address and a
path within the host, in order for the represented resource to be
used.  One way to resolve such identifiers is to use distributed hash
tables~\cite{androutsellis:survey2004,stoica:ton2003}.  Most
distributed hash tables are scalable and are well suited for storing
data indexed by random identifiers.

The most straightforward way to index a resource in a distributed hash
table is to simply store an entry in the table with the resource
identifier as the key and the network location of the resource as the
value.  However, this results in performance and latency issues when a
mobile host contains many resources, since the network location in
each entry must be updated independently whenever the host moves.

For data, this problem can be alleviated by using
replication~\cite{cfs,past}.  However, there are many cases when
replication is not a feasible option, some of which are listed below:

\begin{itemize}
\item Owners of nodes in a distributed hash table may not be willing
  to contribute large amounts of storage to store data for other
  people.  For example, while they may be willing to store network
  locations of home videos, they may not be be willing to store the
  video files themselves.
\item An identifier needs to identify a specific master copy of a file
  in a mobile host in order to ensure that updates to the file are
  immediately available to the host.
\item The resource in question is inherently not replicable.  For
  example, it could be a network service or sensor specific to a
  mobile host.
\end{itemize}

Instead of replication, an alternative approach is to use Mobile
IP~\cite{mobileip} or the Host Identity Protocol~\cite{rfc4423} to
preserve the network address of mobile hosts.  This requires support
in the operating system and the network infrastructure.  Such support
is not wide\-spread, however, so this approach may not be desirable
for applications using distributed hash tables.

In this paper, we propose the use of indirect entries in a distributed
hash table.  An indirect entry contains a host identifier and a
host-specific path.  A host identifier is a random number which
identifies a specific mobile host and is the key to a host entry in
the distributed hash table.  The network address of the mobile host is
obtained from the host entry, which gives the actual network location
when combined with the host-specific path.  When a mobile host moves,
only its host entry needs to be updated.

The remainder of the paper is organized as follows.  We discuss
related work in section~\ref{sec:related-work}.
Section~\ref{sec:batch-updates} describes our proposal and discusses
the circumstances under which it should or should not be used.  We
evaluate it against the straightforward approach in
section~\ref{sec:eval} and conclude in section~\ref{sec:conclude}.

\section{Related work}
\label{sec:related-work}

The use of random numbers for globally unique identifiers is not
uncommon, which takes advantage of the fact that the probability of
two different resources being assigned the same random number is
extremely low for a large enough identifier space.  For example, X.667
defines random-number-based UUIDs~\cite{x667}, while SPKI/SDSI uses
hashes of public keys, which for identification purposes are similar
to random numbers~\cite{rfc2693}.

Ballintijn \etal\ argue that resource naming should be decoupled from
resource identification~\cite{ballintijn:internet2001}.  Resources are
named with human-friendly names, which are based on
DNS~\cite{mockapetris:sigcomm1988}, while identification is done with
object handles, which are globally unique identifiers that need not
contain network locations.  They use DNS to resolve human-friendly
names to object handles and a location service to resolve object
handles to network locations.  The location service uses a
hierarchical architecture for resolving object handles.  This
two-level approach allows the naming of resources without worrying
about replication or migration and the identification of resources
without worrying about naming policies.

Walfish \etal\ argue for the use of semantic-free references for
identifying web documents instead of URLs~\cite{walfish:nsdi2004}.
The reason is that changes in naming policies or ownership of DNS
domain names often result in previous URLs pointing to unrelated or
non-existent documents, even when the original documents still exist.
Semantic-free references are hashes of public keys or other data, and
are resolved to URLs using a distributed hash table based on
Chord~\cite{stoica:ton2003}.  Using semantic-free references would
allow web documents to link to each other without worrying about
changes in the URLs of the documents.

Distributed hash tables, also called peer-to-peer structured overlay
networks, are distributed systems which map a uniform distribution of
identifiers to nodes in the
system~\cite{androutsellis:survey2004,stoica:ton2003,zhao:jsac2004}.
Nodes act as peers, with no node having to play a special role, and a
distributed hash table can continue operation even as nodes join or
leave the system.  Lookups and updates to a distributed hash table are
scalable, typically taking time logarithmic to the number of nodes in
the system.  We experimentally evaluated our work using
OpenDHT~\cite{rhea:sigcomm2005}, which is a public distributed hash
table service based on Bamboo~\cite{bamboo}.

There has also been research on implementing distributed hash tables
on top of mobile ad~hoc
networks~\cite{heer:percomw2006,landsiedel:mobishare2006}.  As with
Mobile IP~\cite{mobileip} and HIP~\cite{rfc4423}, hosts in mobile
ad~hoc networks do not change their network address with movement, so
there would be no need to update entries in a distributed hash table
used for resolving resource identifiers.  However, almost the entire
Internet is not part of a mobile ad~hoc network, so it is of little
help to applications that need to run on current networks.

\section{Batch update for mobile hosts}
\label{sec:batch-updates}

\begin{figure}
  \centering
  \includegraphics[width=7cm]{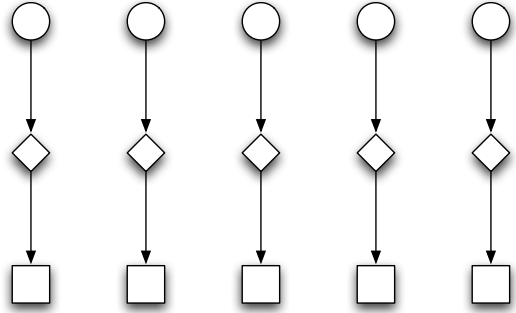}
  \caption{Representation of when identifiers are mapped directly to
    resources in a single mobile host.  Circles denote identifiers,
    diamonds denote direct entries in the hash table, and squares
    denote network locations.  All of the hash table entries must be
    updated whenever the mobile host moves.}
  \label{fig:direct-entries}
\end{figure}

The most straightforward way to map a unique identifier to an actual
network location is to store a \emph{direct entry} in the distributed
hash table for each identifier, with the identifier as the key and the
actual network location as the value.  Resolution is done by simply
looking up the identifier in the distributed hash table and using the
resulting network location.  Figure~\ref{fig:direct-entries}
illustrates this approach.

When the network address of a mobile host changes, this approach
requires that entries for every resource contained by the host be
updated independently.  For a distributed hash table consisting of a
constant number of nodes, this requires time linear to the number of
resources in the mobile host.  When the host contains a large number
of resources, this can result in an unacceptably large delay before
identifiers can be resolved to their updated location.

Instead of storing the network location directly in a hash table entry
for a resource identifier, we propose the alternative approach of
storing both an location-independent identifier for the mobile host
and the host-specific path in an \emph{indirect entry}.  The
\emph{host identifier} identifies the mobile host which contains the
resource and are random numbers as in ordinary resource identifiers.
The distributed hash table contains a host entry which maps this
identifier to the network address of the mobile host.  The path
identifies the specific resource within the host.

In this approach, we first find the corresponding indirect entry for
the given resource identifier in the distributed hash table.  Once the
indirect entry is found, we find the corresponding host entry for the
included host identifier.  We then combine the network address of the
host in the host entry and the path in the indirect entry to construct
the network location of the desired resource.\footnote{When network
  locations are given as a HTTP URL, the network address of the host
  would be an IP address and the port number, while the path would
  simply be the URL path.}  This requires two lookups to the
distributed hash table, compared to a single lookup required for
direct entries.

\begin{figure}
  \centering
  \includegraphics[width=7cm]{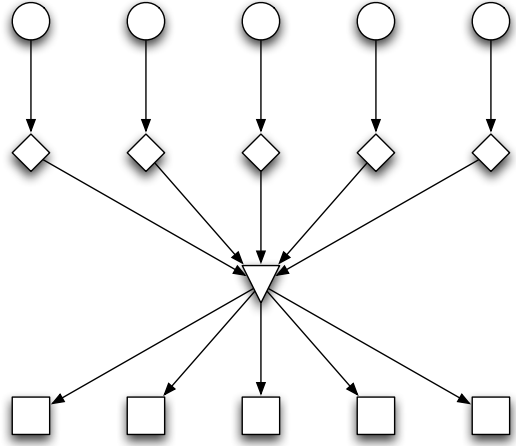}
  \caption{Representation of when identifiers are mapped indirectly to
    resources in a single mobile host.  Circles denote identifiers,
    diamonds denote indirect entries in the hash table, the triangle
    denotes the host entry, and squares denote network locations.
    Only the host entry needs to be updated whenever the mobile host
    moves.}
  \label{fig:indirect-entries}
\end{figure}

However, updating the distributed hash table when a mobile host
changes its network address is much more efficient when using indirect
entries compared to using direct entries.  Unlike with direct entries,
where every entry must be updated independently, only a single host
entry needs to be updated when using indirect entries.  This can
greatly reduce the delay during which resource identifiers cannot be
resolved to their correct network locations.
Figure~\ref{fig:indirect-entries} illustrates the approach using
indirect entries.

\subsection{Using direct and indirect entries together}
\label{sec:concurrent-use}

If a host contains only a very small number of resources or almost
never changes its network address, then using direct entries would be
more efficient because of the smaller lookup overhead.  On the other
hand, using indirect entries drastically reduces the update latency
for a mobile host which contains a large number of resources and changes
its network address frequently.  Fortunately, both types of entries
can be used simultaneously in a single distributed hash table.

Entries in the distributed hash table can be prepended by a magic
number which identifies the type of entry they are.  The magic numbers
are used to distinguish among direct, indirect, and host entries.
They also serve to prevent potential conflicts when the same
distributed hash table is used for other applications besides resource
identifier resolution.  Table~\ref{tab:entries} shows the entry types
and their contents, while figure~\ref{fig:resolution-procedure}
describes the resolution procedure.

\begin{table}
  \centering
  \begin{tabular}{l|l}
    Type & Content \\
    \hline
    Direct entry & MD, network location \\
    Indirect entry & MI, host identifier, path \\
    Host entry & MH, host network address
  \end{tabular}
  \caption{Entry types and their contents.  MD, MI, MH are magic
    numbers for direct, indirect, and host entries, respectively.}
  \label{tab:entries}
\end{table}

\begin{figure}
  \centering
  \framebox{
    \begin{minipage}{0.8\textwidth}
      \begin{enumerate}
      \item Find entry indexed by the resource identifier in the
        distributed hash table.
      \item If entry is direct entry, return with included network
        location.
      \item If entry is indirect entry,
        \begin{enumerate}
        \item Find host entry indexed by the included host identifier.
        \item Combine network address of host in the host entry and
          the path of the resource in the indirect entry to construct
          the network location of the resource.
        \item Return with the network location.
        \end{enumerate}
      \item Otherwise, return that the resource cannot be found.
      \end{enumerate}
    \end{minipage}
  }
  \caption{Resource identifier resolution procedure.  This procedure
    does not treat a host as a resource (extending it so that it does
    is trivial).}
  \label{fig:resolution-procedure}
\end{figure}

\subsection{When to use direct or indirect entries}
\label{sec:threshold}

A host can choose whether to use direct or indirect entries for its
resources depending on which approach performs better for its needs.
But under which circumstances should the host choose which approach?
This section discusses this in terms of lookup overhead and update
latency.

\begin{figure}
  \centering
  \includegraphics[width=4cm]{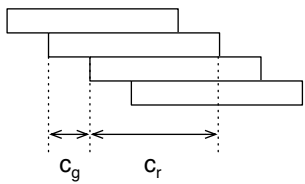}
  \caption{Types of delays during concurrent get operations.  $c_g$ is
    the minimum delay between the operations, while $c_r$ is the delay
    due to network latency.}
  \label{fig:cost-types}
\end{figure}

Since get and put operations to the distributed hash table can be
pipelined, where multiple operations may be handled concurrently as in
figure~\ref{fig:cost-types}, we will consider the time costs $c_g$ or
$c_p$ for an individual get or put operation separately from the fixed
time costs $c_r$ or $c_q$ due to network latency in the get or put
operation, which not only comes from accessing the distributed hash
table externally but also from the communication among the distributed
hash table nodes.  We will assume that the number of nodes in the
distributed hash table is constant so that these costs are also
essentially constant.

When using direct entries, all entries referencing resources in a
given host must be updated independently.  With $n$ resources in a
host, the migration time $c_{m,d}$ required to update all of the
entries when it changes its network address is
\begin{equation}
  \label{eq:direct-migrate}
  c_{m,d} = n c_p + c_q 
\end{equation}

On the other hand, only a single host entry needs to be updated when
using indirect entries, so the migration time $c_{m,i}$ in this case
is
\begin{equation}
  \label{eq:indirect-migrate}
  c_{m,i} = c_p + c_q
\end{equation}

A direct entry requires only a single get operation to resolve an
identifier, whereas an indirect entry requires that it get the
indirect entry first and then obtain the appropriate host entry (the
two get operations cannot be done concurrently since the second
operation is done based on the result from the first operation), so
the respective lookup times $c_{l,d}$ and $c_{l,i}$ are
\begin{eqnarray*}
  c_{l,d} &=& c_g + c_r \\
  c_{l,i} &=& 2 (c_g + c_r)
\end{eqnarray*}

If there are $r_l$ lookups per unit time and $r_m$ migrations per unit
time, then the overall time costs $C_d$ and $C_i$ per unit time when
using direct and indirect entries, respectively, are
\[C_d = r_l c_{l,d} + r_m c_{m,d} = r_l (c_g + c_r) + r_m (n c_p + c_q) \]
\[C_i = r_l c_{l,i} + r_m c_{m,i} = 2 r_l (c_g + c_r) + r_m(c_p + c_q)\]

When minimizing the overall time cost, it is better to use indirect
entries when
\begin{eqnarray*}
  C_i & < & C_d \\
  r_l (c_g + c_r) & < & r_m (n-1) c_p \\
  \frac{r_l}{r_m} & < & \frac{(n-1)c_p}{c_g + c_r}
\end{eqnarray*}

One may also wish to give more weight to reducing migration times or
lookup times.  If we set the weights $w_m$ and $w_l$ by how much
importance we attach to reducing migration times or lookup times,
respectively, we can compute the weighted time costs as
\[C_d' = w_lr_lc_{l,d} + w_mr_mc_{m,d}\]
\[C_i' = w_lr_lc_{l,i} + w_mr_mc_{m,i}\]
and then indirect entries should be used when
\begin{equation}
  \label{eq:threshold}
  \frac{w_l}{w_m} \cdot \frac{r_l}{r_m} < \frac{(n-1)c_p}{c_g+c_r}  
\end{equation}

Assuming a large $n$, with $W$ denoting the relative importance of
reducing lookup times compared to migration times and $R$ denoting how
often lookups occur compared to migrations,
equation~(\ref{eq:threshold}) can be approximately rewritten as
\begin{equation}
  \label{eq:simple-threshold}
  WR < \frac{n c_p}{c_g + c_r}
\end{equation}

Equation~(\ref{eq:simple-threshold}) agrees with our intuition that direct
entries should be used when migration times do not matter or when
migrations are rare, and that indirect entries should be used when
migration times do matter and happen often for mobile hosts with a
large number of resources.  It also gives a concrete forumula for
deciding whether to use direct or indirect entries.

\section{Evaluation}
\label{sec:eval}

\begin{figure}
  \centering
  \resizebox{0.9\textwidth}{!}{\includegraphics{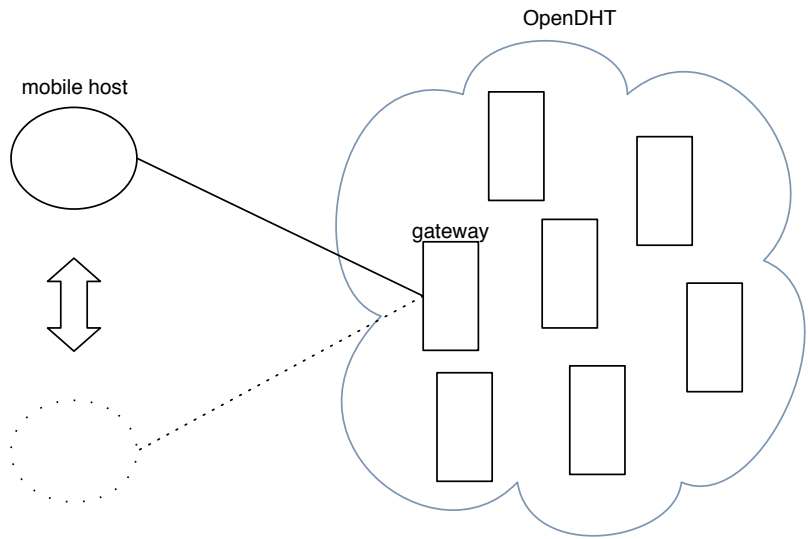}}
  \caption{Experimental setup.  The mobile host, which can move around
    and change its network address, accesses OpenDHT through a gateway
    which is one of the nodes in the distributed hash table.}
  \label{fig:setup}
\end{figure}

In order to evaluate how using direct and indirect entries perform in
a real network, we conducted experiments on
OpenDHT~\cite{rhea:sigcomm2005}.  OpenDHT is a public distributed hash
table service which runs on about 200~nodes in
PlanetLab~\cite{fiuczynski:sigops2006}.  We used the service by
selecting a single gateway to the distributed hash table and accessing
it with the XML-RPC~\cite{xmlrpc} interface throughout our
experiments.  Here the mobile host is not part of the distributed hash
table, similar to how clients are separate from the distributed hash
tables in SFR~\cite{walfish:nsdi2004} and
CoDoNS~\cite{ramasubramanian:sigcomm2004}.\footnote{In cases where
  mobile hosts are part of the distributed hash table, the overhead
  for updating the routing tables should also be considered.}
Figure~\ref{fig:setup} illustrates our experimental setup.

We compared lookup times and migration times when using direct entries
and indirect entries for a single host.  The host, a 2.16~GHz Intel
Core~2 Duo with 1GB of memory connected to the Internet via Ethernet,
was migrated between two network addresses.  Resource identifiers were
mapped to URLs that point to files.  A URL was stored directly in a
direct entry, while only the URL path and a host identifier was stored
in an indirect entry, with the IP address of the host being stored in
a host entry.

\begin{table}
  \centering
  \begin{tabular}{r|r}
    & Lookup time (s) \\ \hline
    Direct & $0.53 \pm 0.63$ \\
    Indirect & $1.13 \pm 0.60$
  \end{tabular}
  \caption{Lookup times and their standard deviations.}
  \label{tab:lookup}
\end{table}

We first measured the lookup times for resolving an identifier to a
URL.  Since lookup for a direct entry requires exactly a single get
operation and lookup for an indirect entry requires exactly two get
operations, lookup times do not depend on the number of resources in a
host.\footnote{The time for a get operation should be constant for a
  distributed hash table with a fixed number of nodes, since network
  latency dominates the time.}  Thus we measured the average lookup
times required by direct and indirect entries by first inserting
entries for 5000 resource identifiers into the distributed hash table
and then resolving randomly selected identifiers 2000 times for each
case.  As expected, the average lookup time for indirect entries was
roughly twice that of direct entries as can be seen in
table~\ref{tab:lookup}.

Next, we measured the migration times when using direct or indirect
entries with varying numbers of resources in the host.  For each
number of resources, we first put in the entries for each resource
into the distributed hash table.  We then migrated the host 100~times
and measured the average migration time.

When updating direct entries, 100~entries were updated concurrently,
which is much faster than updating each entry one by one.  Using
significantly larger amounts of concurrency was problematic because
the gateway to OpenDHT had problems handling the number of
connections.

Also, we selected the entry with the largest remaining time-to-live
value when retrieving entries from OpenDHT.  This entry is the one
that is most up-to-date since we used a fixed TTL value for all
entries.  We did not have to worry about individual entries becoming
large enough to skew the results\footnote{OpenDHT returns all
  unexpired values that have been associated with a key.} since we
alternated the host between only two network addresses.

\begin{figure}
  \centering
  \resizebox{0.9\textwidth}{!}{\includegraphics{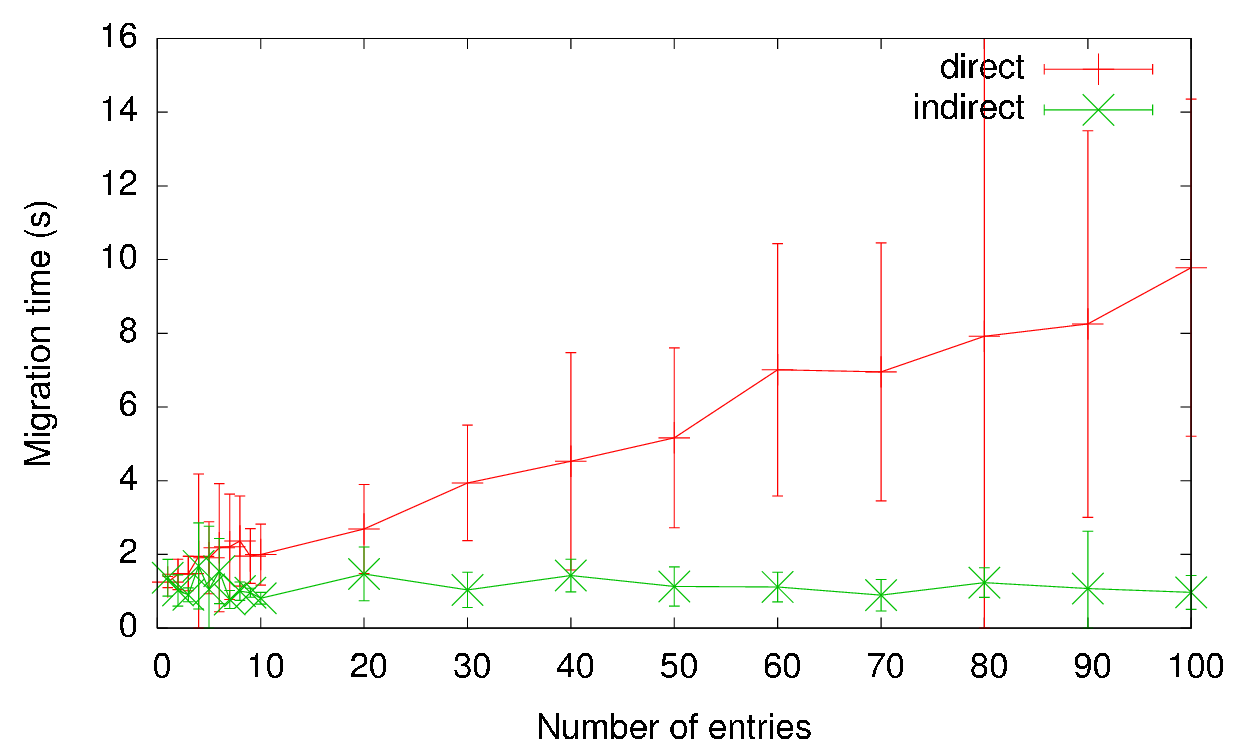}}
  \caption{Migration times for up to 100 resources in the mobile host.
    The error bars denote the standard deviation for each case.}
  \label{fig:migration-100}
\end{figure}

\begin{figure}
  \centering
  \resizebox{0.9\textwidth}{!}{\includegraphics{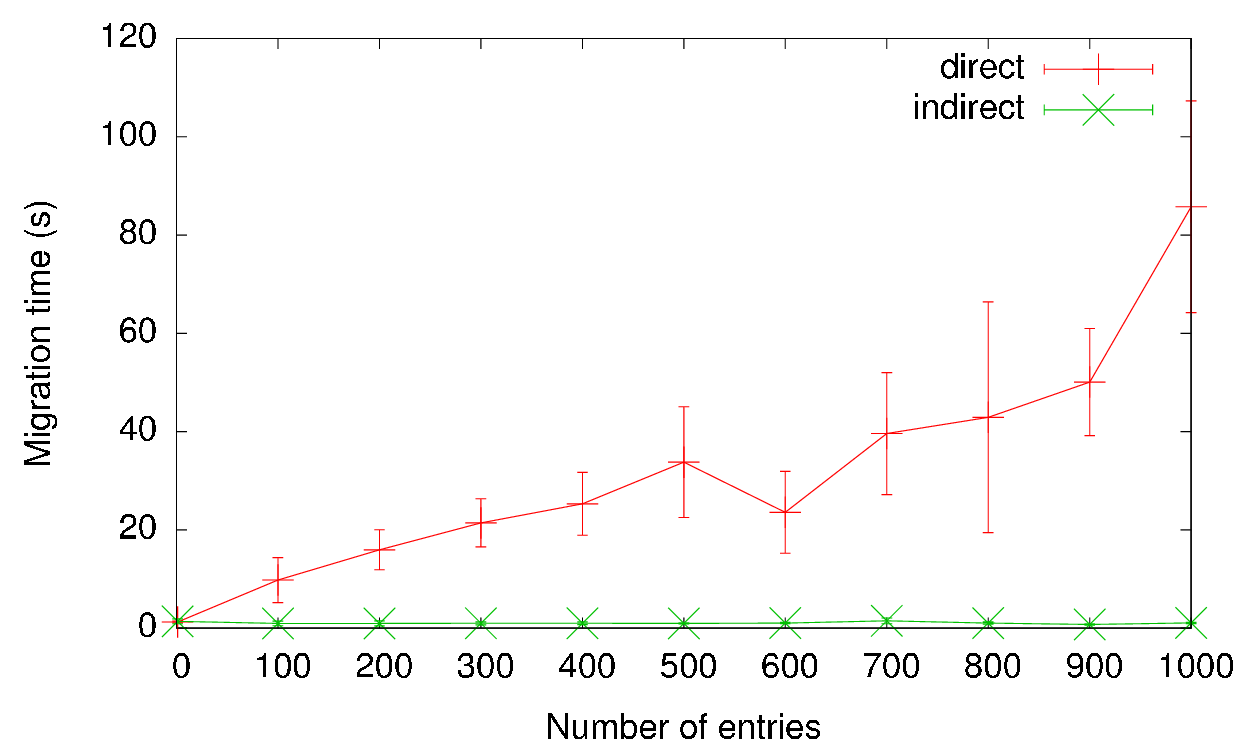}}
  \caption{Migration times for up to 1000 resources in the mobile
    host.  The error bars denote the standard deviation for each
    case.}
  \label{fig:migration-1000}
\end{figure}

\begin{figure}
  \centering
  \resizebox{0.9\textwidth}{!}{\includegraphics{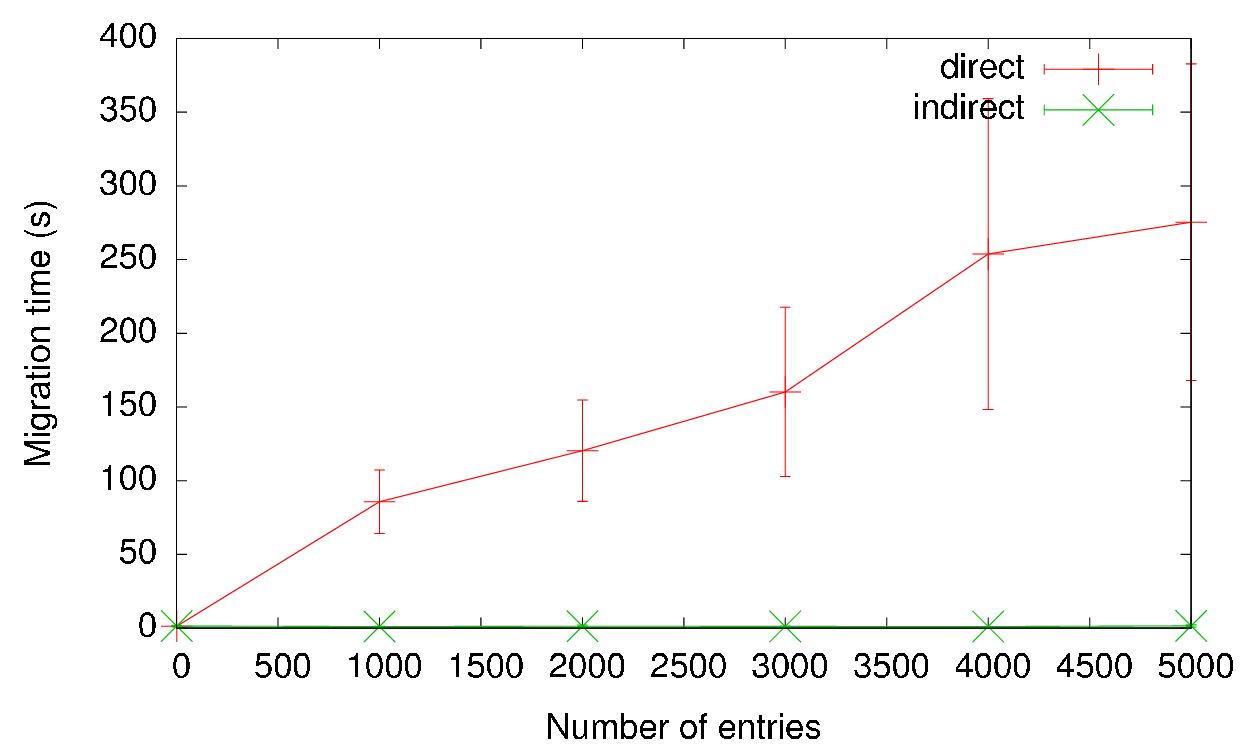}}
  \caption{Migration times for up to 5000 resources in the mobile
    host.  The error bars denote the standard deviation for each
    case.}
  \label{fig:migration-5000}
\end{figure}

Our results for the migration times are shown in figures
\ref{fig:migration-100}, \ref{fig:migration-1000}, and
\ref{fig:migration-5000}, where the host contained up to 100, 1000,
and 5000 resources, respectively.  We can see that migrating direct
entries takes time linear to the number of entries in the host, as is
expected from equation~(\ref{eq:direct-migrate}).  On the other hand,
the time required for updating indirect entries is essentially
constant, as is expected from equation~(\ref{eq:indirect-migrate}).

While migration with indirect entries took only about a second,
migration with direct entries took over 4~minutes and a half with 5000
resources contained in the host.  Even with only 10~resources, it took
about 9~seconds longer to update direct entries compared to the one
second it takes to migrate with indirect entries, which can be a
significant difference for interactive applications.

Mobile hosts could contain even more resources than what was tried in
our experiments.  For example, the home directory in a personal
machine of one of the authors contains more than 60,000 files.  A
straightforward extrapolation from our results suggests that this case
would require almost an hour for migration when using direct entries.

These results show that the drastic reduction in migration time by
using indirect entries over direct entries can be worth the small
increase in lookup time required when resolving indirect entries.

\section{Conclusions}
\label{sec:conclude}

When identifying resources in a distributed system using identifiers
based on random numbers, the most straightforward way to resolve
identifiers with a distributed hash table is to store the network
location directly in the entry keyed by the identifier.  However, when
a mobile host which contains multiple non-replicable resources changes
its network address, all of the associated entries in the distributed
hash table must be updated.

When the number of resources in the mobile host is large, updating all
of the entries so that remote hosts can properly use resources in the
mobile host can take a long time.  Therefore we proposed an
alternative approach, where the entry keyed by the resource identifier
contained only a host identifier and host-specific path for the
resource, and the host identifier itself is a key to a host entry
containing the actual network address for the mobile host.

With our proposed approach, only the host entry needs to be updated
when the mobile host changes its network address.  This can
drastically reduce the delay during which its resources cannot be
resolved to their current network locations, as was shown
theoretically and experimentally.  However, there is a small increase
in the time required for resolving identifiers with our approach, so
we also discussed under which circumstances it should not be used.

In our work, we only consider whether to use direct or indirect
entries given static lookup and migration rates.  It would be
interesting to see how an adaptive system could dynamically adjust the
approach used in order to achieve optimal performance with changing
lookup and migration rates, since such a system would also have to
consider the overhead for switching between the two.

It may also be possible to apply system-specific optimizations to our
approach.  For example, while our approach can be applied to any type
of distributed hash table, it could be possible to reduce the lookup
overhead when it is applied on top of OpenDHT by taking advantage of
the ReDiR framework.

We plan to apply our approach in a decentralized and unified naming
system we are developing, where it would improve performance for
identifying resources such as files and network services located
inside mobile devices in a persistent and location-independent manner.

\bibliography{strings,articles,web,rfc,local,proceedings}

\begin{thebibliography}{10}

\bibitem{androutsellis:survey2004}
Stephanos Androutsellis-Theotokis and Diomidis Spinellis.
\newblock A survey of peer-to-peer content distribution technologies.
\newblock {\em ACM Computing Surveys}, 36(4):335--371, December 2004.

\bibitem{ballintijn:internet2001}
Gerco Ballintijn, Maarten van Steen, and Andrew~S. Tanenbaum.
\newblock Scalable human-friendly resource names.
\newblock {\em IEEE Internet Computing}, 5(5):20--27, September 2001.

\bibitem{w:cooluri}
Tim Berners-Lee.
\newblock Cool {URI}s don't change.
\newblock \url{http://www.w3.org/Provider/Style/URI.html}, 1998.

\bibitem{rfc1738}
Tim Berners-Lee, Larry Masinter, and Mark McCahill.
\newblock {RFC 1738}: Uniform resource locators ({URL}), December 1994.

\bibitem{cfs}
Frank Dabek, M.~Frans Kaashoek, David Karger, Robert Morris, and Ion Stoica.
\newblock Wide-area cooperative storage with {CFS}.
\newblock In {\em Proceedings of the Eighteenth {ACM} Symposium on Operating
  Systems Principles}, pages 202--215, 2001.

\bibitem{rfc2693}
Carl~M. Ellison, Bill Frantz, Butler Lampson, Ron Rivest, Brian Thomas, and
  Tatu Ylonen.
\newblock {RFC 2693}: {SPKI} certificate theory, September 1999.

\bibitem{planetlab}
Marc~E. Fiuczynski.
\newblock {P}lanet{L}ab: Overview, history, and future directions.
\newblock {\em ACM SIGOPS Operating Systems Review}, 40(1):6--10, January 2006.

\bibitem{heer:percomw2006}
Tobias Heer, Stefan G{\"{o}}tz, Simon Rieche, and Klaus Wehrle.
\newblock Adapting distributed hash tables for mobile ad hoc networks.
\newblock In {\em Proceedings of the 4th Annual {IEEE} International Conference
  on Pervasive Computing and Communications Workshops}. IEEE Computer Society
  Press, March 2006.

\bibitem{landsiedel:mobishare2006}
Olaf Landsiedel, Stefan G{\"{o}}tz, and Klaus Wehrle.
\newblock A churn and mobility resistant approach for {DHT}s.
\newblock In {\em Proceedings of the 1st International Workshop on
  Decentralized Resource Sharing in Mobile Computing and Networking}, pages
  42--47. ACM Press, 2006.

\bibitem{mockapetris:sigcomm1988}
Paul~V. Mockapetris and Kevin~J. Dunlap.
\newblock Development of the domain name system.
\newblock {\em ACM SIGCOMM Computer Communication Review}, 18(4):123--133,
  August 1988.

\bibitem{hip}
Robert Moskowitz and Pekka Nikander.
\newblock {RFC 4423}: Host identity protocol ({HIP}) architecture, May 2006.

\bibitem{mobileip}
Charles~E. Perkins.
\newblock {RFC 3344}: {IP} mobility support for {IPv4}, August 2002.

\bibitem{ramasubramanian:sigcomm2004}
Venugopalan Ramasubramanian and Emin~G{\"u}n Sirer.
\newblock The design and implementation of a next generation name service for
  the {I}nternet.
\newblock {\em ACM SIGCOMM Computer Communication Review}, 34(4):331--342,
  October 2004.

\bibitem{bamboo}
Sean Rhea, Dennis Geels, Timothy Roscoe, and John Kubiatowicz.
\newblock Handling churn in a {DHT}.
\newblock In {\em Proceedings of the {USENIX} Annual Technical Conference},
  June 2004.

\bibitem{rhea:sigcomm2005}
Sean Rhea, Brighten Godfrey, Brad Karp, John Kubiatowicz, Sylvia Ratnasamy,
  Scott Shenker, Ion Stoica, and Harlan Yu.
\newblock {OpenDHT}: A public {DHT} service and its uses.
\newblock In {\em Proceedings of the 2005 Conference on Applications,
  Technologies, Architectures, and Protocols for Computer Communications},
  pages 73--84. ACM Press, 2005.

\bibitem{past}
Antony Rowstron and Peter Druschel.
\newblock Storage management and caching in {PAST}, a large-scale, persistent
  peer-to-peer storage utility.
\newblock In {\em Proceedings of the Eighteenth {ACM} Symposium on Operating
  Systems Principles}, pages 188--201, 2001.

\bibitem{stoica:ton2003}
Ion Stoica, Robert Morris, David Liben-Nowell, David~R. Karger, M.~Frans
  Kaashoek, Frank Dabek, and Hari Balakrishnan.
\newblock {C}hord: A scalable peer-to-peer lookup protocol for internet
  applications.
\newblock {\em IEEE/ACM Transactions on Networking}, 11(1):17--32, February
  2003.

\bibitem{steen:comm1998}
Maarten van Steen, Franz~J. Hauck, Philip Homburg, and Andrew~S. Tanenbaum.
\newblock Locating objects in wide-area systems.
\newblock {\em IEEE Communications Magazine}, 36(1):104--109, January 1998.

\bibitem{walfish:nsdi2004}
Michael Walfish, Hari Balakrishnan, and Scott Shenker.
\newblock Untangling the {W}eb from {DNS}.
\newblock In {\em Proceedings of the First Symposium on Networked Systems
  Design and Implementation}, pages 225--238, March 2004.

\bibitem{xmlrpc}
Dave Winer.
\newblock {XML-RPC} specification.
\newblock \url{http://www.xmlrpc.com/spec}, June 1999.

\bibitem{x667}
Procedures for the operation of {OSI} registration authorities: Generation and
  registration of universally unique identifiers ({UUID}s) and their use as
  {ASN.1} object identifier components.
\newblock ITU-T Recommendation X.667, September 2004.

\bibitem{zhao:jsac2004}
Ben~Y. Zhao, Ling Huang, Jeremy Stribling, Sean~C. Rhea, Anthony~D. Joseph, and
  John~D. Kubiatowicz.
\newblock {T}apestry: A resilient global-scale overlay for service deployment.
\newblock {\em IEEE Journal on Selected Areas in Communications}, 22(1):41--53,
  January 2004.

\end{thebibliography}
\bibliographystyle{plain}

\end{document}